\documentclass[referee]{mn2e}
\usepackage{tipa}
\usepackage{epsfig}
\usepackage{color}

\begin{document}
\title[Self-similar flows with magnetic fields]{Self-similar structure of magnetized ADAFs and CDAFs}
\author[Dong Zhang and Z. G. Dai]{Dong Zhang\thanks{dongzhanghz@gmail.com}
and Z.G. Dai\thanks{dzg@nju.edu.cn (ZGD)}
\\Department of Astronomy, Nanjing University, Nanjing 210093,
China } \maketitle

\begin{abstract}
We study the effects of a global magnetic field on
viscously-rotating and vertically-integrated accretion disks around
compact objects using a self-similar treatment. We extend Akizuki \&
Fukue's work (2006) by discussing a general magnetic field with
three components ($r, \varphi, z$) in advection-dominated accretion
flows (ADAFs). We also investigate the effects of a global magnetic
field on flows with convection. For these purposes, we first adopt a
simple form of the kinematic viscosity $\nu=\alpha
c_{s}^{2}/\Omega_{K}$ to study magnetized ADAFs: a vertical and
strong magnetic field, for instance, not only prevents the disk from
being accreted but also decreases the isothermal sound speed. Then
we consider a more realistic model of the kinematic viscosity
$\nu=\alpha c_{s}H$, which makes the infall velocity increase but
the sound speed and toroidal velocity decrease. We next use two
methods to study magnetized flows with convection, i.e., we take the
convective coefficient $\alpha_{c}$ as a free parameter to discuss
the effects of convection for simplicity. We establish the
$\alpha_{c}-\alpha$ relation for magnetized flows using the
mixing-length theory and compare this relation with the
non-magnetized case. If $\alpha_{c}$ is set as a free parameter,
then $|v_{r}|$ and $c_{s}$ increase for a large toroidal magnetic
field, while $|v_{r}|$ decreases but $|v_{\varphi}|$ increases (or
decreases) for a strong and dominated radial (or vertical) magnetic
field with increasing $\alpha_{c}$. In addition, the magnetic field
makes the $\alpha_{c}-\alpha$ relation be distinct from that of
non-magnetized flows, and allows the $\rho\propto r^{-1}$ or
$\rho\propto r^{-2}$ structure for magnetized non-accreting
convection-dominated accretion flows with $\alpha+g\alpha_{c}< 0$
(where $g$ is the parameter to determine the condition of convective
angular momentum transport).
\end{abstract}
\begin{keywords}
accretion, accretion disks --- black hole physics --- MHD
\end{keywords}

\section{Introduction}
Rotating accretion flows with viscosity and angular momentum
transfer can be divided into several classes, depending on different
structures and energy transfer mechanisms in the flows:
advection-dominated accretion flows (ADAFs), advection-dominated
inflow-outflows (ADIOs), convection-dominated accretion flows
(CDAFs), neutrino-dominated accretion flows (NDAFs) and
magnetically-dominated accretion flows (MDAFs).

ADAFs were introduced by Ichimaru (1977) and then have been widely
studied over thirty years. The opically-thick ADAFs with
super-Eddington accretion rates were discussed by Abramowicz et al.
(1988) in details (see also Begelman et al. 1982; Eggum et al.
1988). The optically-thin ADAFs with low, sub-Eddington accretion
rates were discussed by Rees et al. (1982) and Narayan \& Yi (1994,
1995a, 1995b) (see also Ichimaru 1977; Abramowicz et al. 1995;
Gammie \& Popham 1998; Popham \& Gammie 1998; Wang \& Zhou 1999). In
particular, Narayan \& Yi (1994) introduced self-similar solutions
for ADAFs with the fixed ratio of the advective cooling rate to the
viscous heating rate in the disk. Wang \& Zhou (1999) solved
self-similar solutions for optically-thick ADAFs. The effects of
general relativity were considered in Gammie \& Popham (1998) and
Popham \& Gammie (1998).

CDAFs were presented in details in Narayan et al. (2000, hereafter
NIA). They discussed the effects of convection on angular momentum
and energy transport, and presented the relations between the
convective coefficient $\alpha_{c}$ and the classical viscosity
parameter $\alpha$. A non-accreting solution can be obtained when
convection moves angular momentum inward and the viscosity parameter
$\alpha$ is small. Later, a series of works have been published to
discuss the disk structure, the MHD instability, the condition of
angular momentum transport in CDAFs (e.g., Igumenshchev et al. 2000,
2002, 2003; Quataert \& Gruzinov 2000; Narayan et al. 2002;
Igumenshchev 2002; Lu et al. 2004; van der Swaluw et al. 2005).

The effects of a magnetic field on the disk were also studied (see
Balbus \& Hawley 1998; Kaburiki 2000; Shadmehri 2004; Meier 2005;
Shadmehri \& Khajenabi 2005, 2006; Akizuki \& Fukue 2006; Ghanbari
et al. 2007). Balbus \& Hawley (1998) discussed the MHD turbulence
initiated by magnetorotational instability (MRI) and its effects on
the angular momentum transportation. Kaburaki (2000) considered an
analytic model to describe the ADAFs with a global magnetic field
and Meier (2005) considered how a turbulent and magnetized disk
creates a global well-ordered magnetic field, and introduced a
magnetically-dominated flow. Shadmehri (2004) and Chanbari et al.
(2007) discussed the self-similar structure of the magnetized ADAFs
in spherical polar coordinates. Moreover, Shamehri \& Khajenabi
(2005, 2006, hereafter SK05, SK06) presented self-similar solutions
of flows based on the vertically integrated equations. They
discussed the relations between magnetic fields components in
different directions, and mainly focused on the effects of the
magnetic field on the disk structure. Akizuki \& Fukue (2006,
hereafter AF06), different from SK05 and SK06, emphasized an
intermediate case where the magnetic force is comparable to other
forces by assuming the physical variables in the disk only as
functions of radius. However, they merely discussed a global
toroidal magnetic field in the disk.

In this paper, we first extend the work of AF06 by considering a
general large-scale magnetic field in all the three components in
cylindrical coordinates ($r, \varphi, z$) and then discuss effects
of the global magnetic field on the flows with convection. We adopt
the treatment that the flow variables are functions of the disk
radius, neglect the different structure in the vertical direction
except for the $z$-component momentum equation. We also discuss
magnetized accretion flows with convection, and compare our results
with those in NIA, in which a large-scale magnetic field is
neglected.

This paper is organized as follows: basic equations are presented in
\S2. We obtain self-similar solutions in \S3 and discuss the effects
of a general large-scale magnetic field on the disk flow. In \S4 we
investigate the structure and physical variables in magnetized
CDAFs, and present the relation of the convective parameter
$\alpha_{c}$ and the classical viscosity parameter $\alpha$. We
adopt a more realistic form of the kinematic viscosity in \S5. Our
conclusions are presented in \S6.

\section{Basic Equations}

In this paper, we use all quantities with their usual meanings: $r$
is the radius of the disk, $v_{r}$ and $v_{\varphi}$ are the radial
and rotation velocity, $\Omega=v_{\varphi}/r$ is the angular
velocity of the disk, $\Omega_{K}=(GM/r^{3})^{1/2}$ is the Keplerian
angular velocity, $\Sigma=2\rho H$ is the disk surface density with
$\rho$ to be the disk density and $H$ to be the half-thickness, and
$c_{s}=(p/\rho)^{1/2}$ is the isothermal sound speed with $p$ to be
the gas pressure in the disk.

Moreover, we consider a magnetic field in the disk with three
components $B_{r}$, $B_{\varphi}$ and $B_{z}$ in the cylindrical
coordinates ($r, \varphi, z$). We define the Alfv\'{e}n sound speeds
$c_{r}$, $c_{\varphi}$ and $c_{z}$ in three directions of the
cylindrical coordinates as
$c_{r,\varphi,z}^{2}=B_{r,\varphi,z}^{2}/(4\pi \rho)$. We consider
that all flow variables are only functions of radius $r$, and write
basic equations, i.e., the continuity equation, the three components
($r,\varphi,z$) of the momentum equation and the energy equation:

\begin{equation}
\frac{1}{r}\frac{d}{dr}\left(r\Sigma
v_{r}\right)=2\dot{\rho}H,\label{e10}
\end{equation}

\begin{equation}
v_{r}\frac{dv_{r}}{dr}=\frac{v_{\varphi}^{2}}{r}-\frac{GM}{r^{2}}-\frac{1}{\Sigma}\frac{d}{dr}\left(\Sigma
c_{s}^{2}\right)-\frac{1}{2\Sigma}\frac{d}{dr}\left(\Sigma
c_{z}^{2}+\Sigma
c_{\varphi}^{2}\right)-\frac{c_{\varphi}^{2}}{r},\label{e11}
\end{equation}

\begin{equation}
\frac{v_{r}}{r}\frac{d(rv_{\varphi})}{dr}=\frac{1}{\Sigma
r^{2}}\frac{d}{dr}\left(\Sigma \alpha
\frac{c_{s}^{2}}{\Omega_{K}}r^{3}\frac{d\Omega}{dr}\right)+\frac{c_{\varphi}c_{r}}{r}
+\frac{c_{r}}{\sqrt{\Sigma}}\frac{d}{dr}\left(\sqrt{\Sigma
}c_{\varphi}\right),\label{e12}
\end{equation}

\begin{equation}
\Omega_{K}^{2}H-\frac{1}{\sqrt{\Sigma}}c_{r}\frac{d}{dr}\left(\sqrt{\Sigma}
c_{z}\right)=\frac{c_{s}^{2}+\frac{1}{2}\left(c_{\varphi}^{2}+c_{r}^{2}\right)}{H},\label{e13}
\end{equation}

\begin{equation}
\frac{v_{r}}{\gamma-1}\frac{dc_{s}^{2}}{dr}-v_{r}\frac{c_{s}^{2}}{\rho}\frac{d\rho}{dr}=f\frac{\alpha
c_{s}^{2}r^{2}}{\Omega_{K}}\left(\frac{d\Omega}{dr}\right)^{2}.\label{e131}
\end{equation}
Here we consider the height-integrated equations using the classical
$\alpha$-prescription model with $\alpha$ to be the viscosity
parameter, and use the Newtonian gravitational potential. In the
mass continuity equation, we also consider the mass loss term
$\partial \rho/\partial t$. In the energy equation we take $\gamma$
to be the adiabatic index of the disk gas and $f$ to measure the
degree to which the flow is advection-dominated (NY94), and neglect
the Joule heating rate.

In AF06, a general case of viscosity
$\eta=\rho\nu=\Omega_{K}^{-1}\alpha p_{\rm gas}^{\mu}(p_{\rm
gas}+p_{\rm mas})^{1-\mu}$ with $\mu$ to be a parameter is
mentioned. If the ratio of the magnetic pressure to the gas pressure
is constant (as assumed in the self-similar structure), the solution
of the basic equations can be obtained with replacing $\alpha$ by
$\alpha(1+\beta)^{1-\mu}$. In our paper, however, we first adopt the
classical form $\nu=\alpha c_{s}^{2}/\Omega_{K}$ for simplicity in
\S 3 and \S 4, in which we mainly focus on the effects of a magnetic
field on the variables $v_{r}$, $v_{\varphi}$ and $c_{s}$. A more
realistic model requires $\nu=\alpha c_{s}H$ with both $c_{s}$ and
$H$ as functions of the magnetic field strength. We discuss this
model in \S 5 and compare it with the results in \S 3 and \S 4.

Our equations are somewhat different from those in SK05 and SK06,
since we only consider the disk variables as functions of radius
$r$, while SK05 and SK06 discuss the magnetic field structure in the
vertical direction. More details about the basic equations are
discussed in Appendix A. When $B_{r}=0$ and $B_{z}=0$, our equations
switch back to the equations in AF06, in which only the toroidal
magnetic field is considered and all the variables are taken to
depend merely on radius $r$.

In addition, we need the the three-component induction equations to
measure the magnetic field escaping rate:
\begin{equation}
\dot{B_{r}}\approx 0,\label{e14}
\end{equation}

\begin{equation}
\dot{B_{\varphi}}=\frac{d}{dr}\left(v_{\varphi}B_{r}-v_{r}B_{\varphi}\right),\label{e15}
\end{equation}

\begin{equation}
\dot{B_{z}}=-\frac{d}{dr}\left(v_{r}B_{z}\right)-\frac{v_{r}B_{z}}{r}.\label{e16}
\end{equation}

\section{Self-Similar Solutions for ADAF}
If we assume the parameters $\gamma$ and $f$ in the energy equation
are independent of radius $r$, then we can adopt a self-similar
treatment similar to NY94 and AF06,
\begin{equation}
v_{r}(r)=-c_{1}\alpha\sqrt{\frac{GM}{r}},\label{e201}
\end{equation}

\begin{equation}
v_{\varphi}(r)=c_{2}\sqrt{\frac{GM}{r}},\label{e202}
\end{equation}

\begin{equation}
c_{s}^{2}(r)=c_{3}\frac{GM}{r},\label{e203}
\end{equation}

\begin{equation}
c_{r,\varphi,z}^{2}(r)=\frac{B_{r,\varphi,z}^{2}}{4\pi
\rho}=2\beta_{r,\varphi,z}c_{3}\frac{GM}{r},\label{e204}
\end{equation}
where the coefficients $c_{1}$, $c_{2}$ and $c_{3}$ are similar to
those in AF06, and $\beta_{r}$, $\beta_{\varphi}$ and $\beta_{z}$
measure the ratio of the magnetic pressure in three directions to
the gas pressure, i.e., $\beta_{r,\varphi,z}=p_{{\rm
mag},r,\varphi,z}/p_{\rm gas}$. Following AF06, we also denote the
structure of the surface density $\Sigma$ by
\begin{equation}
\Sigma(r)=\Sigma_{0}r^{s}.\label{e205}
\end{equation}
The half-thickness of the disk still satisfies the relation
$H\propto r$ and we obtain
\begin{equation}
H(r)=H_{0}r.\label{e206}
\end{equation}

Substituting self-similar relations (\ref{e201})--(\ref{e206}) to
equations (\ref{e11}), (\ref{e12}) and (\ref{e131}), we can obtain
the algebraic equations of $c_{1}$, $c_{2}$ and $c_{3}$:
\begin{equation}
-\frac{1}{2}c_{1}^{2}\alpha^{2}=c_{2}^{2}-1-[(s-1)+\beta_{z}(s-1)+\beta_{\varphi}(s+1)]c_{3},\label{e207}
\end{equation}

\begin{equation}
-\frac{1}{2}c_{1}c_{2}\alpha=-\frac{3}{2}\alpha(s+1)c_{2}c_{3}+c_{3}(s+1)\sqrt{\beta_{r}\beta_{\varphi}},\label{e208}
\end{equation}

\begin{equation}
c_{2}^{2}=\frac{4}{9f}\left(\frac{1}{\gamma-1}+s-1\right)c_{1}.\label{e209}
\end{equation}

If in the cylindrical coordinates we assume the three components of
magnetic field $B_{r,\varphi,z}>0$, then $v_{\varphi}(r)$ can be
either positive or negative, depending on the detailed magnetic
field structure in the disk. In a particular case where $B_{r}=0$ or
$B_{\varphi}=0$, we can only obtain the value of $|c_{2}|$, but in a
general case where $B_{r}B_{\varphi}\neq0$, we are able to determine
the value of $c_{2}$.

Figures 1 and 2 show the self-similar coefficients $c_{1}$,
$|c_{2}|$ and $c_{3}$ as functions of the advection parameter $f$
with different $(\beta_{r}, \beta_{\varphi}, \beta_{z})$. We
consider the disk to be radiation dominated with $\gamma=4/3$, and
take $s=-1/2$ (i.e., $\rho\propto r^{-3/2}$ as the common case) and
the viscosity parameter $\alpha=0.1$, which is the widely used
value.

Figure 1 shows changes of the coefficients $c_{1}$, $|c_{2}|$,
$c_{3}$ with $\beta_{z}$ and $\beta_{\varphi}$. We neglect the
radial magnetic fields $B_{r}$, and take the parameters $(\beta_{r},
\beta_{\varphi}, \beta_{z})$ in the left three panels in Figure 1
for (0, 1, 0), (0, 1, 1), (0, 1, 3) and (0, 1, 10), and then take
the value of $\beta_{\varphi}$ to be 10 in the right three panels.
As $\beta_{r}=0$, we can only obtain $|c_{2}|$ without needing to
determine the direction of $v_{\varphi}$. The coefficients $c_{1}$
and $c_{3}$ increase with increasing the advection parameter $f$,
but $|c_{2}|$ decreases monotonously as a function of $f$ except for
a strong toroidal magnetic field. Moreover, with the fixed ratio
$\beta_{z}$, an increase of $\beta_{\varphi}$ makes all the
coefficients $|c_{i}|$ become larger. Oppositely, $|c_{i}|$
decreases with increasing $\beta_{z}$. In fact, with a small radial
magnetic field $\beta_{r}\approx 0$, we can obtain an analytical
solution of $c_{i}$ from equations (\ref{e207})-(\ref{e209}), which
are similar to expressions (27)-(29) in AF06, but we should replace
$(1-s)/(1+s)$ by $(1-s)(1+\beta_{z})/(1+s)$ and $\beta$ by
$\beta_{\varphi}$ in those expressions instead. Also, we have
$c_{2}^{2}\propto f^{-1}c_{1}$, $c_{3}\propto c_{1}$, and
$c_{1}\propto \beta_{\varphi}$ for large $\beta_{\varphi}$ and
$c_{1}\propto \beta_{z}^{-1}$ for large $\beta_{z}$, all of which
are consistent with the results in Figure 1.

As a result, from Figure 1, we first find that a strong toroidal
magnetic field leads to an increase of the infall velocity
$|v_{r}|$, rotation velocity $|v_{\varphi}|$ and isothermal sound
speed $c_{s}$, and $|v_{r}|$ and $c_{s}$ are large in the case where
the disk flow is mainly advection-dominated, but the rotation
velocity $|v_{\varphi}|$ increases with increasing $f$ only in the
case where the toroidal magnetic field is large enough. This
conclusion is consistent with the case 1 in AF06. Second, the high
ratio $\beta_{z}$ decreases the value of $|v_{r}|$, $|v_{\varphi}|$
and $c_{s}$, which means that a strong magnetic pressure in the
vertical direction prevents the disk matter from being accreted, and
decreases the effect of gas pressure as accretion proceeds.

Figure 2 shows how $c_{1}$, $|c_{2}|$ and $c_{3}$ change with
$\beta_{r}$ and $\beta_{\varphi}$, where we neglect the vertical
magnetic $B_{z}$. For a small value of $\beta_{r}$, the coefficients
$c_{1}$, $|c_{2}|$ and $c_{3}$ also increase with increasing
$\beta_{\varphi}$. However, a change of $c_{i}$ is not obvious for a
large value of $\beta_{r}$. We are able to calculate the limiting
value of $c_{i}$ in the extreme case where $\beta_{r}$ is large
enough and $\beta_{r}\beta_{\varphi}\neq 0$ using an analytical
method. From equations (\ref{e207})-(\ref{e209}), we can obtain
$c_{1}=2/[\epsilon''+\sqrt{(\epsilon'')^{2}+2\alpha^{2}}]$ and
$c_{2}^{2}=\epsilon''c_{1}$ for large $\beta_{r}$, where
$\epsilon''=\frac{4}{9f}\left\{(\gamma-1)^{-1}+s-1\right\}$. If
$\epsilon''\gg \alpha$, we have $c_{1}\sim (\epsilon'')^{-1}\propto
f$ and $|c_{2}|\sim 1$, which means that the infall velocity
$|v_{r}|$ increases with advection parameter $f$ linearly, and the
radial velocity $|v_{\varphi}|$ is nearly the Keplerian velocity, no
matter whether the disk is efficiently cooled or not. Also, for a
large value of $\beta_{r}$, equation (\ref{e208}) becomes
$-(c_{1}c_{2}\alpha)/2\sim
c_{3}(s+1)\sqrt{\beta_{r}\beta_{\varphi}}$. Since $c_{1},c_{3}>0$ in
the accretion disk, we obtain $c_{2}<0$, which means that the
direction of rotation in the disk is opposite to the toroidal
magnetic field $B_{\varphi}$. Actually, in the case where
$\beta_{r}$ is sufficient large and $\beta_{r}\beta_{\varphi}\neq
0$, the angular momentum transported due to the magnetic field
stress is dominated over that due to the viscosity, and balances
with the advection angular momentum. As we take $B_{r,\varphi,z}>0$,
from equation (\ref{e12}), we obtain that the large angular momentum
due to the magnetic field stress makes the value of the advection
angular momentum ($\dot{M}v_{\varphi}r$, where $\dot{M}$ is the mass
accretion rate) increase in the disk, which requires $v_{\varphi}<0$
in the self-similar structure\footnote{In some previous works (e.g.,
Wang 95, Lai 98, SK05 and SK06), the rotation velocity is taken to
be positive and the toroidal magnetic field $B_{\varphi}$ to be
negative. The advection transports angular momentum inward, while
the magnetic stress transports angular momentum outward instead.
This previous result is consistent with our result here if we change
the cylindrical coordinate used above from ($r,\varphi,z$) to
($r,-\varphi,z$). In this paper it is convenient for us to take
$B_{r,\varphi,z}>0$ and to obtain s series of self-similar solutions
about magnetized flows in many different cases.}.

From the mass-continuity equation (\ref{e10}) and the induction
equations (\ref{e14})-(\ref{e16}) as well as the solved coefficients
$c_{i}$, we can solve the self-similar structure of the mass loss
and magnetic field escaping rate with forms of
$\dot{\rho}=\dot{\rho}_{0}r^{s-5/2}$ and
$\dot{B}_{r,\varphi,z}=\dot{B}_{r_0,\varphi_0,z_0}r^{(s-5)/2}$ where
$\dot{\rho}_{0}$ satisfies

\begin{equation}
\dot{\rho}_{0}=-\left(s+\frac{1}{2}\right)\frac{c_{1}\alpha\Sigma_{0}\sqrt{GM}}{2H_{0}}.\label{e210}
\end{equation}
As mentioned in AF06, when $s=-1/2$, i.e., $\Sigma\propto r^{-1/2}$
or $\rho\propto r^{-3/2}$, there is no wind in the disk, and thus we
can use the formula $-2\pi rv_{r}\Sigma =\dot{M}$ to determine the
surface density $\Sigma$. In addition, in the region of the disk
where is adiabatic with $\rho \propto r^{-1/(\gamma-1)},\,\,
p\propto r^{-\gamma/(\gamma-1)},\,\, v_r \propto
r^{(3-2\gamma)/(\gamma-1)}$ (i.e., $s=(\gamma-2)/(\gamma-1)$), we
obtain the self-similar solution of $c_{1}\alpha=\sqrt{2}, c_{2}=0$
and $c_{3}=0$, which describe the Bondi accretion. However, if the
disk region satisfies the entropy-conservation condition with $f=0$,
we can still obtain an accretion-disk solution beyond the
self-similar treatment. For a CDAF with $\rho\propto r^{-1/2}$
(NIA), a steady disk without wind requires $c_{1}=0$ or $v_{r}=0$.

$\dot{B}_{r_0,\varphi_0,z_0}$ satisfy
\begin{equation}
\dot{B}_{r_0}\approx 0,\label{e211}
\end{equation}

\begin{equation}
\dot{B}_{\varphi_0}=\left(\frac{s-3}{2}\right)GM
\left\{c_{2}\sqrt{\frac{4\pi\beta_{r}c_{3}\Sigma_{0}}{H_{0}}}+c_{1}
\alpha\sqrt{\frac{4\pi\beta_{\varphi}c_{3}\Sigma_{0}}{H_{0}}}\right\},\label{e2111}
\end{equation}

\begin{equation}
\dot{B}_{z_0}=\left(\frac{s-1}{2}\right)c_{1}\alpha(GM)\sqrt{\frac{4\pi\beta_{z}\Sigma_{0}c_{3}}{H_{0}}},\label{e212}
\end{equation}
where $H_{0}$ in the expression (\ref{e206}) can be obtained from
the hydrostatic equilibrium equation (\ref{e13}), that is,
\begin{equation}
H_{0}=\frac{1}{2}\left[(s-1)c_{3}\sqrt{\beta_{r}\beta_{z}}+\sqrt{c_{3}^{2}(s-1)^{2}\beta_{r}\beta_{z}
+4(1+\beta_{\varphi}+\beta_{r})c_{3}}\right],\label{e213}
\end{equation}
and thus we obtain the half-thickness of the disk
$H=H_{0}c_{s}/(\sqrt{c_{3}}\Omega_{K})$. We will discuss the effects
of the magnetic field on $H$ later in \S 5.

\section{Self-Similar Solutions for CDAF}

In CADFs, both advection and convection play contributions to the
angular momentum and energy transportation. We propose a CDAF model
in a global magnetic field in order to compare it with
non-globally-magnetized CDAFs. We follow the idea of NIA in this
section and consider the effect of a magnetic field\footnote{We
adopt the ($\alpha, \alpha_{c}$)-prescription following NIA. The MHD
simulations beyond this prescription can be seen in Igumenshchev et
al. (2002, 2003), Hawley \& Balbus (2002) and so on. Moreover,
Quataert \& Gruzinov (2000) also develop an analytical model for
CDAFs. }. The viscosity angular momentum flux is
\begin{equation}
\dot{J}_{v}=-\alpha\frac{c_{s}^{2}}{\Omega_{K}}\rho
r^{3}\frac{d\Omega}{dr},\label{e301}
\end{equation}
and the convection angular momentum flux can be written as
\begin{equation}
\dot{J}_{c}=-\alpha_{c}\frac{c_{s}^{2}}{\Omega_{K}}\rho
r^{3(1+g)/2}\frac{d}{dr}\left(\Omega
r^{3(1-g)/2}\right),\label{e3011}
\end{equation}
where $\alpha_{c}$ is the dimensionless coefficient to measure the
strength of convective diffusion, $g$ is the parameter to determine
the condition of convective angular momentum transport. Convection
transports angular momentum inward (or outward) for $g<0$ (or $>0$).

The energy equation of a CADF is
\begin{equation}
\rho
v_{r}T\frac{ds}{dr}+\frac{1}{r^{2}}\frac{d}{dr}\left(r^{2}F_{c}\right)=Q^{+}=f\frac{(\alpha+g\alpha_{c})
\rho
c_{s}^{2}r^{2}}{\Omega_{K}}\left(\frac{d\Omega}{dr}\right)^{2},\label{e302}
\end{equation}
where the convective energy flux $F_{c}$ is
\begin{equation}
F_{c}=-\alpha_{c}\frac{c_{s}^{2}}{\Omega_{K}}\rho
T\frac{ds}{dr},\label{e303}
\end{equation}
where we still consider the general energy equation without vertical
integration as in NIA, and still neglect the Joule heating rate.

Using the angular momentum equation, the energy equation of the CDAF
and the self-similar structure (\ref{e201})-(\ref{e206}), we can
obtain the relevant algebraic equations of self-similar structure of
the CDAF,
\begin{equation}
-\frac{1}{2}c_{1}c_{2}\alpha=-\frac{3}{2}(\alpha+g\alpha_{c})(s+1)c_{2}c_{3}
+c_{3}(s+1)\sqrt{\beta_{r}\beta_{\varphi}},\label{e304}
\end{equation}

\begin{equation}
\left(s-\frac{1}{2}\right)\left(\frac{1}{\gamma-1}+s-1\right)c_{3}\alpha_{c}+\left(\frac{1}{\gamma-1}+s-1\right)c_{1}\alpha
=(\alpha+g\alpha_{c})\frac{9f}{4}c_{2}^{2}.\label{e305}
\end{equation}
The radial momentum equation is still the same as that in ADAF.
Combining equation (\ref{e207}), (\ref{e304}) and (\ref{e305}), we
can finally solve the coefficients $c_{1}$, $c_{2}$, and $c_{3}$ in
the case of CDAF and compare them with those in ADAF. The
dimensionless coefficient $\alpha_{c}$ can be calculated using the
mixing length theory, and we adopt equation (15) in NIA, who
describes the relation of $\alpha_{c}$ with $s$, $c_{3}$ and
$\gamma$ (in NIA, they used the symbols $a$ and $c_{0}$, where
$a=1-s$, and $c_{0}^{2}=c_{3}$ in our paper).

To simplify the problem, we first prefer using a simpler treatment
with a fixed $\alpha_{c}$ to discussing the solutions of
(\ref{e207}), (\ref{e304}) and (\ref{e305}), i.e., we take
$\alpha_{c}$ as a free parameter rather than a calculated variable,
since $\alpha_{c}$ does not dramatically change in many cases. Then
we can adopt a similar treatment as in \S3 to solve equations
(\ref{e207}), (\ref{e304}) and (\ref{e305}). Similarly as in \S 3,
we first use the analytical method to discuss some particular cases.

When $\beta_{r}\beta_{\varphi}\sim 0$ (which implies that the radial
or toroidal magnetic fields are weak), we can obtain an analytical
solution similar to that in \S3 (see Appendix B for more details
about the calculation). We discuss two cases. One is that the
toroidal magnetic field $B_{\varphi}$ is dominated and the radial
magnetic field is weak ($B_{r}\approx 0$). Then we can have an
approximate solution,
\begin{equation}
c_{1}\alpha\sim \frac{2\beta_{\varphi}}{3(\alpha+g\alpha_{c})},
\label{e306}
\end{equation}

\begin{equation}
c_{2}^{2}=
\frac{c_{1}\alpha}{\alpha+g\alpha_{c}}\left[\epsilon''-|\xi''|\frac{\alpha_{c}}{3(\alpha+g\alpha_{c})(s+1)}\right],
\label{e3071}
\end{equation}

\begin{equation}
c_{3}\sim \frac{2\beta_{\varphi}}{9(\alpha+g\alpha_{c})^{2}(s+1)}.
\label{e3072}
\end{equation}
where $\xi''=\frac{4}{9f}(s-\frac{1}{2})(\frac{1}{\gamma-1}+s-1)$.
From these equations, we know the coefficients $c_{1}$ and $c_{3}$
increase with increasing the convective parameter $\alpha_{c}$ for
$g<0$ (i.e., convection transports angular momentum inward), but we
cannot obtain the relation between $\alpha_{c}$ and $|c_{2}|$ unless
the values of $s$, $g$ and $\gamma$ are given in detail. The other
case is that the vertical magnetic field $B_{z}$ is dominated. In
this case, we obtain
\begin{equation}
c_{1}\alpha\sim \frac{3(\alpha+g\alpha_{c})(1+s)}{(1-s)\beta_{z}},
\label{e3073}
\end{equation}

\begin{equation}
c_{2}^{2}=
\frac{3(1+s)}{(1-s)\beta_{z}}\left[\epsilon''-|\xi''|\frac{\alpha_{c}}{3(\alpha+g\alpha_{c})(s+1)}\right],
\label{e3074}
\end{equation}

\begin{equation}
c_{3}\sim \frac{1}{(1-s)\beta_{z}}. \label{e3072}
\end{equation}
We find that the coefficients $c_{1}$ and $|c_{2}|$ decrease with
increasing $\alpha_{c}$ for $g<0$ while the value of $c_{3}$ is more
or less the same for a fixed $\beta_{z}$.

Another analytical solution can be obtained when $\beta_{r}$ is
large and $\beta_{\varphi}\neq 0$, and we have the relations
\begin{equation}
c_{1}\propto
\left[\frac{\alpha}{\alpha+g\alpha_{c}}\epsilon''+\sqrt{\left(\frac{\alpha}{\alpha+g\alpha_{c}}\right)^{2}\epsilon''^{2}
+2\alpha^{2}}\right]^{-1},\label{e308}
\end{equation}

\begin{equation}
|c_{2}|\propto
\left[\epsilon''+\sqrt{\epsilon''^{2}+2(\alpha+g\alpha_{c})^{2}}\right]^{-1/2},
\label{e309}
\end{equation}

\begin{equation}
c_{3}\propto -c_{1}c_{2}. \label{e3091}
\end{equation}
From formulae (\ref{e308}) and (\ref{e309}),  we again find changes
of $c_{1}$ and $|c_{2}|$ with $\alpha_{c}$, which is similar to the
former case. From equation (\ref{e3091}) and $c_{1,3}>0$, we get
$c_{2}<0$, which has also been obtained in \S 3.

Figure 3 shows some examples of the effect of the convection
parameter $\alpha_{c}$ on the three coefficients $c_{i}$. In order
to see the results clearly, we take $\alpha=1$ and change the value
of $\alpha_{c}$ from 0 to 0.9 with several sets of magnetic field
parameters ($\beta_{r}$, $\beta_{\varphi}$, $\beta_{z}$)$=(0,3,0)$,
$(3,3,0)$ and $(0,0,3)$. Also we set $\gamma=4/3$, $s=-1/2$ and
$g=-1/3$. The basic results in Figure 3 are consistent with the
above discussion using the analytical method. In particular, we
notice that the three coefficients do not change dramatically in the
case of $\beta_{r}\sim \beta_{\varphi}$, since the magnetic field
gives a contribution to the angular momentum rather than the
viscosity, and reduces the effect of convection on the disk.

Next we want to obtain the relation between $\alpha_{c}$ and
$\alpha$ following NIA, i.e., we consider $\alpha_{c}$ to be the
variable as a function of $s$, $\gamma$ and $c_{3}$. Using the
treatment in NIA based on the mixing length theory and equations
(\ref{e207}), (\ref{e304}) and (\ref{e305}), we can establish the
$\alpha_{c}$-$\alpha$ relation. NIA discussed such a relation with
$g=1$ and $g=-1/3$, and find that the solution with $s=-1/2$ is
available only for $\alpha$ greater than a certain critical
$\alpha_{\rm crit}$ when the isothermal sound speed reaches its
maximum value, and the value of $\alpha_{c}$ decreases monotonously
as $\alpha$ increases. However, our results are quite different from
those in NIA for two reasons. First, we keep the term
$v_{r}dv_{r}/dr$ in the radial momentum equation, while in NIA this
term is neglected. As a result, in many cases, the sound speed to
determine the actual critical $\alpha$ for available solutions does
not reach exactly its maximum value. Second and more importantly, we
consider the effect of the large-scale magnetic field on the disk.

From equation (\ref{e207}), we obtain
\begin{equation}
\frac{1}{2}c_{1}^{2}\alpha^{2}+[(1-s)(1+\beta_{z})-\beta_{\varphi}(1+s)]c_{3}-1
<0.\label{e3092}
\end{equation}
If the radial magnetic field is weak, then we have
$c_{3}<[(1-s)\beta_{z}]^{-1}$ for a large vertical magnetic field
and $c_{3}<2\beta_{\varphi}/[9(\alpha+g\alpha_{c})^{2}(1+s)]$ for a
large toroidal magnetic field. On the other hand, from NIA, we have
$c_{3}>\gamma/[(2-s)(2+s\gamma-s)]$ for the convective process to be
available. Therefore, the structure of flows with convection cannot
be maintained for a large vertical magnetic field. Moreover, if the
term $\beta_{r}\beta_{\varphi}$ is large, we still obtain a small
value of $c_{3}\sim
c_{1}|c_{2}|\alpha/(2\sqrt{\beta_{r}\beta_{\varphi}})$ with a small
value of $\alpha_{c}$, which is almost independent of the variation
of $\alpha$.

Figure 4 shows examples of the $\alpha_{c}-\alpha$ relation with
different magnetic field structures. We take $\gamma=1.4$ and
$s=-1/2$. The left panel shows the $\alpha_{c}-\alpha$ relation with
different values of $\beta_{\varphi}$. When the magnetic field is
small ($\beta_{\varphi}$=0 and 1 in this panel), $\alpha_{c}$
decreases with increasing $\alpha$, and $\alpha$ has its critical
(minimum) value for the solution to be available. These results are
basically consistent with those in NIA. However, when
$\beta_{\varphi}$ becomes large, $\alpha_{c}$ increases as the
viscosity parameter $\alpha$ increases, and the critical value of
$\alpha$ becomes extremely small or even disappears. The right panel
of Figure 4 shows the $\alpha_{c}-\alpha$ relation with different
values of $\beta_{z}$. When $\beta_{z}$ becomes large, the critical
value of $\alpha$ also disappears, but $\alpha$ has its maximum
value. This result is quite different from NIA, who found that only
the minimum value of $\alpha$ exists and becomes important.

From the above discussion, we conclude that a strong vertical
magnetic field or large $\beta_{r}\beta_{\varphi}$ prevents the
convective process in flows, while a moderate vertical magnetic
field is available for small $\alpha$. A strong toroidal magnetic
field with weak radial field makes the convective process become
important even for large $\alpha$ in flows.

In NIA, a self-similar convection-driven non-accreting solution with
$s=1/2$ (i.e. $\Sigma\propto r^{1/2}$) was given for
$\alpha+g\alpha_{c}=0$ when $\alpha$ is smaller than the critical
value $\alpha_{\rm crit}$. However, the relation
$\alpha+g\alpha_{c}=0$ cannot be satisfied if
$\beta_{r}\beta_{\varphi}\neq 0$ for magnetized CDAFs. In fact, we
are still able to get a self-similar structure for magnetized CDAFs
when the $\alpha_{c}-\alpha$ relation mentioned is no longer
satisfied (i.e., inequality (\ref{e3092}) is not satisfied). For
$\beta_{r}\beta_{\varphi}\neq 0$, the zero infall velocity (i.e.
$c_{1}=0$) requires $s=-1$ ($\rho\propto r^{-2}$) from equation
(\ref{e304}), and $\alpha$ as a function of $c_{3}$,
\begin{equation}
\alpha=\alpha_{c}\left(|g|-\frac{|\xi''|c_{3}}
{c_{2}^{2}}\right)<\alpha_{c}|g|,\label{e311}
\end{equation}
with the maximum value of $\alpha_{c}$ to be
\begin{equation}
\alpha_{c,\rm
crit2}=\frac{(1+\beta_{z})}{9\sqrt{2}}\sqrt{\frac{9-(2\beta_{z}+5)\gamma}{2\gamma(1+\beta_{z})}}.\label{e312}
\end{equation}

Furthermore, if we turn the radial momentum equation from its
vertical integration to its general from, we can still have the
relation (\ref{e311}) but $s=0$ ($\rho\propto r^{-1}$), and the
maximum value of $\alpha_{c}$ to be
\begin{equation}
\alpha_{c,\rm
crit2}=\frac{(1+\beta_{z})}{4\sqrt{2}}\sqrt{\frac{2-(1+\beta_{z})\gamma}{\gamma(1+\beta_{z})}}.\label{e313}
\end{equation}
Such a structure of $\rho\propto r^{-1}$ was also obtained by
Igumenshchev et al. (2003), who explained the structure as a result
of vertical leakage of convective energy flux from the disk. In our
model, however, we show that this structure is due to the
inefficient angular momentum transfer by viscosity and the zero
Lorentz force in the $\varphi$-direction.

As a result, we obtained a self-similar solution for magnetized
CDAFs with $c_{1}=0$, $s=-1$ or $s=0$ (for the general form) and
$\alpha+g\alpha_{c}< 0$. This solution is adopted when the normal
self-similar solutions mentioned above for convective flows cannot
be satisfied.

\section{A more realistic form of kinematic viscosity}
In the above sections \S3 and \S4, we assume the kinematic viscosity
$\nu=\alpha c_{s}^{2}/\Omega_{K}$ and take the viscosity parameter
$\alpha$ as a constant in our discussion for simplicity. A more
realistic model based on the physical meaning of the viscosity
parameter is $\nu=\alpha c_{s}H$ with $H\neq c_{s}/\Omega_{K}$ in
the magnetized disk. In this section we consider the effect of
different forms of kinematic viscosity $\nu$. In order to compare
with the results in \S3 and \S4, we replace $\alpha$ in the last two
sections by $\alpha'$ and take $\nu=\alpha
c_{s}H=\alpha'c_{s}^{2}/\Omega_{K}$ in this section. Also, we still
adopt the definition of $c_{1}$ using equation (\ref{e201}). From
formula (\ref{e213}), we are able to obtain
\begin{equation}
\alpha'=\alpha\left\{\left[\left(\frac{1-s}{2}\right)^{2}\beta_{r}\beta_{z}c_{3}
+(1+\beta_{r}+\beta_{\varphi})\right]^{1/2}
-\left(\frac{1-s}{2}\right)(\beta_{r}\beta_{z}c_{3})^{1/2}\right\}.\label{e601}
\end{equation}
When $\beta_{r,z}=0$, we have
$\alpha'=\alpha\sqrt{1+\beta_{\varphi}}$ and equation (\ref{e601})
switches back to the case of $\mu=1/2$ in AF06. For large
$\beta_{r}$ or $\beta_{\varphi}$ and small $\beta_{z}$, we have
$\alpha'\sim \alpha\sqrt{1+\beta_{r}+\beta_{\varphi}}$ and
$\alpha'\gg \alpha$. For large $\beta_{z}$, we obtain $\alpha'\sim
\alpha(1+\beta_{r}+\beta_{\varphi})/(1-s)\sqrt{\beta_{r}\beta_{z}c_{3}}$
and $\alpha'\ll \alpha$. This result can be explained as being due
to the fact that a large toroidal or radial magnetic field makes the
disk half-thickness $H$ become large and increase the kinematic
viscosity (since $\nu\propto H$), but a large vertical field reduces
the height $H$ and decreases the kinematic viscosity.

Figure 5 shows the effect of a modified kinematic viscosity on the
three coefficients $c_{1}$, $|c_{2}|$ and $c_{3}$ in ADAFs. A more
realistic expression of $\nu$ increases the infall velocity, but
decreases the radial velocity and the isothermal sound speed.
However, a difference between these two cases of kinematic viscosity
is obvious for a large toroidal magnetic field rather than a large
vertical field. In fact, if the toroidal magnetic field
$B_{\varphi}$ is strong and dominated in ($B_{r}, B_{\varphi},
B_{z}$), we can adopt a similar solution of (49)-(51) in AF06 for
$\mu=1/2$, and find that $c_{1}$ increases but $|c_{2}|$ and $c_{3}$
reaches their limiting values with increasing $\beta_{\varphi}$. If
the radial magnetic field $B_{r}$ is strong and dominated, we can
obtain $c_{1}\sim {\rm const}$, $|c_{2}|\propto \beta_{r}^{-1/4}$
and $c_{3}\propto \beta_{r}^{-1/2}$, which are different from \S3 in
which $|c_{2}|\sim 1$ for large $\beta_{r}$. Furthermore, if
$\beta_{z}$ is large enough, we have the limiting value $c_{1} \sim
3(1+\beta_{r}+\beta_{\varphi})(1-s)^{-3/2}\beta_{z}^{-1}\beta_{r}^{-1/2}$,
$c_{3}\sim (1-s)^{-1}\beta_{z}^{-1}$ and
$c_{2}^{2}=3\epsilon''(s+1)c_{3}$, and the values of $|c_{2}|$ and
$c_{3}$ are more or less the same, no matter what the form of
kinematic viscosity is.

For flows with convection, it is convenient for us to adopt the
general definition of $\alpha_{c}$ from NIA, which measures a degree
of convection in the flows. We find that the conclusions in \S4 are
not basically changed if we replace $\alpha$ in \S4 by $\alpha'$.
The $\alpha'-\alpha_{c}$ relation can be turned back to the
$\alpha-\alpha_{c}$ relation using equation (\ref{e601}). However,
there is no dramatic change between these two relations except for
extremely strong magnetic fields.

\section{Conclusions}
In this paper we have studied the effects of a global magnetic field
on viscously-rotating and vertically-integrated accretion disks
around compact objects using a self-similar treatment. Our
conclusions are listed as follows:

(1) We have extended Akizuki and Fukue's self-similar solutions
(2006) by considering a three-component magnetic field $B_{r}$,
$B_{\varphi}$, and $B_{z}$ in ADAFs. If we set the kinematic
viscosity $\nu=\alpha c_{s}^{2}/\Omega_{K}$ as its classical form,
then with the flow to be advection-dominated, the infall velocity
$|v_{r}|$ and the isothermal sound speed $c_{s}$ increase, and even
the radial velocity $|v_{\varphi}|$ can exceed the Keplerian
velocity with a strong toroidal magnetic field. The strong magnetic
field in the vertical direction prevents the disk from being
accreted, and decreases the effect of the gas pressure. For a large
radial magnetic field, $v_{r}$, $v_{\varphi}$ and $c_{s}$ can reach
their limiting values, and the direction of radial velocity is
actually negative, since the angular momentum transfer due to the
magnetic field stress in this case is dominated over that due to the
viscosity in the disk, and makes the value of advection angular
momentum increase inward.

(2) If the convective coefficient $\alpha_{c}$ in flows is set as a
free parameter, $|v_{r}|$ and $c_{s}$ increase with increasing
$\alpha_{c}$ for large $B_{\varphi}$ and weak $B_{r}$. Also,
$|v_{r}|$ becomes smaller and $|v_{\varphi}|$ becomes larger (or
smaller) with increasing $\alpha_{c}$ for a strong and dominated
radial (or vertical) magnetic field.

(3) The $\alpha_{c}-\alpha$ relation in the magnetized disk is
different from that in the non-magnetized disk. For large
$B_{\varphi}$ and weak $B_{r}$, $\alpha_{c}$ increases with
increasing $\alpha$, the critical value $\alpha_{\rm crit}$ to
determine different cases of the $\alpha_{c}-\alpha$ relation
disappears, and $\Sigma\propto r^{-1/2}$ can be satisfied for any
value of $\alpha$. A moderate vertical magnetic field is available
for small $\alpha$. The large $B_{z}$ or $B_{r}B_{\varphi}$, on the
other hand, prevents the convective process in flows.

(4) The self-similar convection envelope solution in NIA should be
replaced by $c_{1}=0$, $\alpha+g\alpha_{c}<0$ and $s=-1$
($\rho\propto r^{-2}$) for the vertical integration form of angular
equations and  $s=0$ ($\rho\propto r^{-1}$) for the general form in
magnetized CDAFs. This solution can be adopted in the region that
does not satisfy the normal self-similar solutions for flows with
convection and $\alpha_{c}<\alpha_{c,\rm crit2}$.

(5) The magnetic field increases the disk height $H$ for large
$B_{r}$ and $B_{\varphi}$, but decreases it for large $B_{z}$ in the
magnetized disk. A more realistic model of the kinematic viscosity
$\nu=\alpha c_{s}H$ makes the infall velocity in ADAFs increase and
the sound speed and toroidal velocity decrease compared with the
simple case when the form $\nu=\alpha c_{s}^{2}/\Omega_{K}$ is
assumed.

\section*{Acknowledgements}
We would like to thank the referee, Jun Fukue, for useful comments.
We also thank X. D. Li and Y. W. Yu for their helpful discussions.
This work is supported by the National Natural Science Foundation of
China (grants 10221001 and 10640420144) and the National Basic
Research Program of China (973 program) No. 2007CB815404.

\begin{appendix}
\section{}
The momentum equation of accretion flows can be written as (Frank et
al. 2002)
\begin{equation}
(\textbf{v}\cdot\nabla)\textbf{v} =-\frac{1}{\rho}\nabla
P-\nabla\Phi+\Omega^{2}\textbf{r}+(\nabla\cdot\sigma)+\frac{1}{\rho
c}\textbf{j}\times \textbf{B},\label{e401}
\end{equation}
where $\sigma$ is the viscosity stress tensor, $\textbf{j}\times
\textbf{B}/(\rho c)$ is the density Lorentz force. Also, the
Amp\`{e}re's law and the induction equation (Faraday's law) are
\begin{equation}
\textbf{j}=\frac{c}{4\pi}\left(\nabla\times
\textbf{B}\right).\label{e402}
\end{equation}

\begin{equation}
\frac{\partial \textbf{B}}{\partial t}=\nabla\times(\textbf{v}\times
\textbf{B})+\eta_{m}\nabla^{2}\textbf{B}.\label{e403}
\end{equation}
where $\eta_{m}=c^{2}/(4\pi\sigma_{e})$ is the magnetic diffusivity
and $\sigma_{e}$ is the electrical conductivity. For simplicity, we
consider the extreme case that $\sigma_{e}\rightarrow \infty$ and
$\eta_{m}\approx 0$ , and then neglect the second term in the right
side of the induction equation (\ref{e403}). Combining equations
(\ref{e401}) and (\ref{e402}), we can obtain the three components of
the momentum equation. In particular, the three components of the
Lorentz force in the cylindrical coordinates are
\begin{equation}
\frac{4\pi}{c}(\textbf{j}\times
\textbf{B})_{r}=-\frac{1}{2}\frac{\partial}{\partial
r}(B_{z}^{2}+B_{\varphi}^{2})+B_{z}\frac{\partial B_{r}}{\partial
z}-\frac{B_{\varphi}^{2}}{r},\label{e404}
\end{equation}

\begin{equation}
\frac{4\pi}{c}(\textbf{j}\times
\textbf{B})_{\varphi}=\frac{1}{r}B_{\varphi}B_{r}+B_{r}\frac{\partial
B_{\varphi}}{\partial r}+B_{z}\frac{\partial B_{\varphi}}{\partial
z},\label{e405}
\end{equation}

\begin{equation}
\frac{4\pi}{c}(\textbf{j}\times
\textbf{B})_{z}=-\frac{1}{2}\frac{\partial}{\partial
z}(B_{r}^{2}+B_{\varphi}^{2})+B_{r}\frac{\partial B_{z}}{\partial
r}.\label{e406}
\end{equation}

Based on the consideration that all flow variables including the
magnetic field are mainly functions of radius $r$, we can conclude
$v_{z}=0$ and $\partial/\partial z=0$. Or a more realistic
consideration requires $\partial/\partial z\sim (H/r)
\partial/\partial r\ll \partial/\partial r$. Also, we take $\partial/\partial \varphi=0$ for the axisymmetric disk.
We rewrite the Lorentz force using the Alfven sound speed as
\begin{equation}
\frac{1}{\rho c}(\textbf{j}\times
\textbf{B})_{r}=-\frac{1}{2\rho}\frac{\partial}{\partial
r}[\rho(c_{z}^{2}+c_{\varphi}^{2})]-\frac{c_{\varphi}^{2}}{r},\label{e407}
\end{equation}

\begin{equation}
\frac{1}{\rho c}(\textbf{j}\times
\textbf{B})_{\varphi}=\frac{1}{r}c_{\varphi}c_{r}+\frac{c_{r}}{\sqrt{\rho}}\frac{\partial}{\partial
r}(\sqrt{\rho}c_{\varphi}),\label{e408}
\end{equation}

\begin{equation}
\frac{1}{\rho c}(\textbf{j}\times
\textbf{B})_{z}=\frac{c_{r}}{\sqrt{\rho}}\frac{\partial}{\partial
r}(\sqrt{\rho}c_{z}),\label{e409}
\end{equation}

These expressions are different from SK05 and SK06, who considered
the magnetic field structure as a function of both radius $r$ and
height $z$: $B_{r}(r,z)=z(B_{r})_{H}/H$,
$B_{\varphi}(r,z)=z(B_{\varphi})_{H}/H$ with $H$ to be the
half-thickness of the disk, and $B_{z}(r,z)=B_{z}(r)$. However, in
this paper we take the magnetic field to be homogeneous in the
vertical direction and neglect the term of $\partial/\partial z$ as
mentioned above except for the $z$-component equation, in which we
take the total pressure as $P_{\rm tot}=P_{\rm
gas}+(B_{\varphi}^{2}+B_{r}^{2})/8\pi$ in the vertical direction,
and adopt $\partial P_{\rm tot}/\partial z\sim-P_{\rm tot}/H$ to
estimate the value of $H$. In \S2, we use the height-integration
equations.

\section{}
Expressions (29)-(37) in \S4 can be derived as follows:

When $\beta_{r}=0$ or $\beta_{\varphi}=0$, we obtain equations for
the three coefficients $c_{i}$ in CDAFs as
\begin{equation}
-\frac{1}{2}c_{1}^{2}\alpha^{2}=c_{2}^{2}-1-[(s-1)(1+\beta_{z})+(1+s)\beta_{\varphi}]c_{3},\label{e501}
\end{equation}

\begin{equation}
c_{1}\alpha=3(\alpha+g\alpha_{c})(s+1)c_{3},\label{e502}
\end{equation}

\begin{equation}
c_{2}^{2}=\epsilon''\frac{\alpha
c_{1}}{\alpha+g\alpha_{c}}+\xi''\frac{\alpha_{c}c_{3}}{\alpha+g\alpha_{c}},\label{e503}
\end{equation}
with $\epsilon''=\frac{4}{9f}(\frac{1}{\gamma-1}+s-1)$ and
$\xi''=\frac{4}{9f}(s-\frac{1}{2})(\frac{1}{\gamma-1}+s-1)$. Then we
can write the equation for $c_{1}$ as
\begin{equation}
\frac{1}{2}c_{1}^{2}\alpha^{2}+c_{1}\alpha\left\{\epsilon''
+\xi''\frac{\alpha_{c}}{3(s+1)(\alpha+g\alpha_{c})}
+\frac{1}{3}\left[\left(\frac{1-s}{1+s}\right)(1+\beta_{z})-\beta_{\varphi}\right]\right\}-1=0.\label{e504}
\end{equation}
When $\beta_{z}$ is large, the above equation can be simplified as
\begin{equation}
\frac{1}{2}c_{1}^{2}\alpha^{2}+\frac{c_{1}\alpha}{3}\left(\frac{1-s}{1+s}\right)
\frac{\beta_{z}}{\alpha+g\alpha_{c}}-1=0,\label{e505}
\end{equation}
and we obtain
\begin{equation}
c_{1}\alpha\sim
\frac{3(\alpha+g\alpha_{c})(1+s)}{(1-s)\beta_{z}}\label{e506}.
\end{equation}
Similarly, we can get the solution for large $\beta_{\varphi}$ and
small $\beta_{r}$.

On the other hand, if the radial magnetic field is strong and
dominated and $\beta_{\varphi}\neq0$, then equation (\ref{e502})
should be replaced by
\begin{equation}
-\frac{1}{2}c_{1}c_{2}\alpha=(s+1)c_{3}\sqrt{\beta_{r}\beta_{\varphi}},\label{e507}
\end{equation}
and we obtain an equation in the extreme case,
\begin{equation}
\frac{1}{2}c_{1}^{2}\alpha^{2}+\epsilon''\frac{\alpha
c_{1}}{\alpha+g\alpha_{c}}-1=0\label{e508},
\end{equation}
and get
\begin{equation}
\alpha
c_{1}=2(\alpha+g\alpha_{c})\left[\epsilon''+\sqrt{\epsilon''^{2}+2(\alpha+g\alpha_{c})^{2}}\right]^{-1}.\label{e509}
\end{equation}

\end{appendix}

\newpage
\begin{figure}
\resizebox{\hsize}{!} {\includegraphics{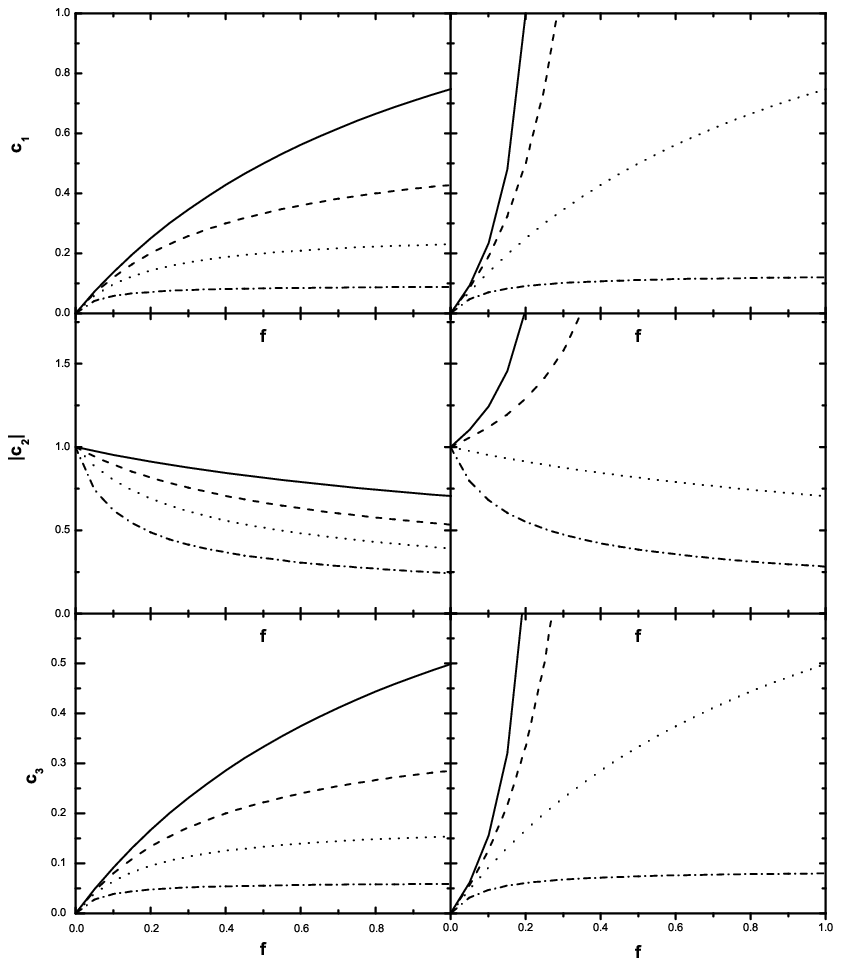}} \caption{The
self-similar coefficients $c_{1}$, $|c_{2}|$, and $c_{3}$ as
functions of the advection parameter $f$ for different sets of
parameters $\beta_{r}$, $\beta_{\varphi}$ and $\beta_{z}$. We take
$\alpha$=0.1, $\gamma=4/3$ and $s=-1/2$. The left three panels
correspond to ($\beta_{r}$, $\beta_{\varphi}$ $\beta_{z}$)$=(0,1,0)$
({\em solid lines}), $(0,1,1)$ ({\em dashed lines}), $(0,1,3)$ ({\em
dotted lines}), and (0,1,10) ({\em dash-dotted lines}). The right
three panels correspond to ($\beta_{r}$, $\beta_{\varphi}$
$\beta_{z}$)$=(0,10,0)$ ({\em solid lines}), $(0,10,1)$ ({\em dashed
lines}), $(0,10,3)$ ({\em dotted lines}), and $(0,10,10)$ ({\em
dash-dotted lines}).}
\end{figure}

\newpage
\begin{figure}
\resizebox{\hsize}{!} {\includegraphics{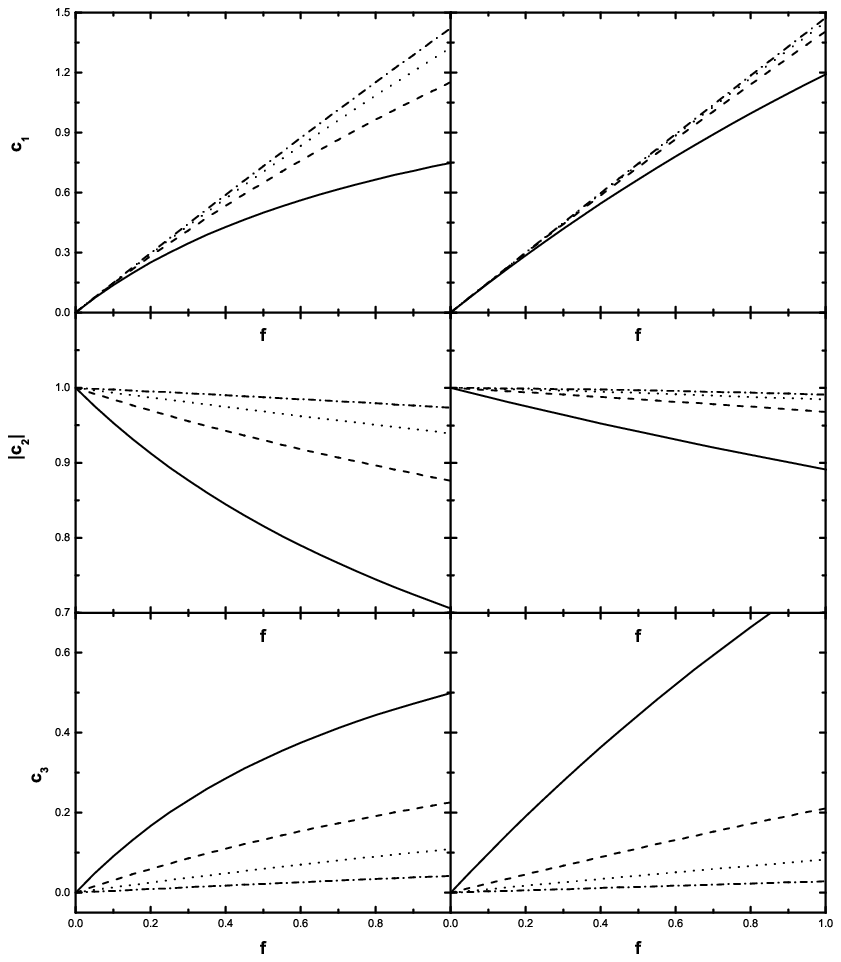}} \caption{The
self-similar coefficients $c_{1}$, $|c_{2}|$, and $c_{3}$ as
functions of the advection parameter $f$ for different sets of
parameters $\beta_{r}$, $\beta_{\varphi}$ and $\beta_{z}$ with
$\alpha$=0.1, $\gamma=4/3$ and $s=-1/2$. The left three panels
correspond to ($\beta_{r}$, $\beta_{\varphi}$ $\beta_{z}$)$=(0,1,0)$
({\em solid lines}), $(0.1,1,0)$ ({\em dashed lines}), $(1,1,0)$
({\em dotted lines}), and $(10,1,0)$ ({\em dash-dotted lines}). The
right three panels correspond to ($\beta_{r}$, $\beta_{\varphi}$
$\beta_{z}$)=$(0,2.5,0)$ ({\em solid lines}), $(0.1,2.5,0)$ ({\em
dashed lines}), $(1,2.5,0)$ ({\em dotted lines}), and $(10,2.5,0)$
({\em dash-dotted lines}).}
\end{figure}

\newpage
\begin{figure}
\resizebox{\hsize}{!} {\includegraphics{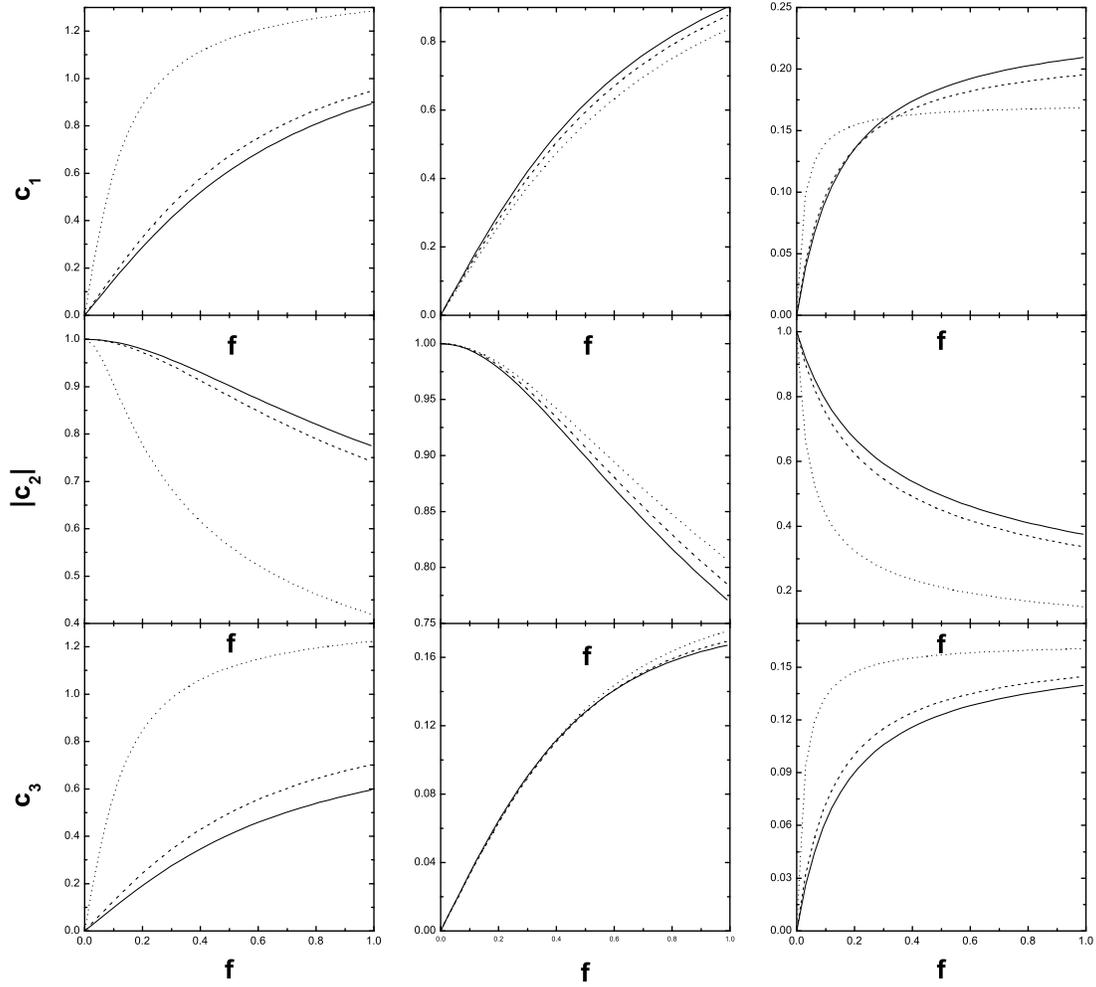}} \caption{The
coefficients $c_{1}$, $|c_{2}|$, and $c_{3}$ as functions of $f$
with different sets of parameters $\alpha_{c}$ and ($\beta_{r}$,
$\beta_{\varphi}$ $\beta_{z}$). We take $\alpha$=1, $\gamma=4/3$,
$s=-1/2$ and $g=-1/3$. The left three panels correspond to
($\beta_{r}$, $\beta_{\varphi}$ $\beta_{z}$)$=(0,3,0)$, the middle
three panels to ($\beta_{r}$, $\beta_{\varphi}$
$\beta_{z}$)$=(3,3,0)$, and the right panels to ($\beta_{r}$,
$\beta_{\varphi}$ $\beta_{z}$)$=(0,0,3)$. Different lines refer to
$\alpha_{c}=0$ ({\em solid lines}), $\alpha_{c}=0.3$ ({\em dashed
lines}) and $\alpha_{c}=0.9$ ({\em dotted lines}).}
\end{figure}

\newpage
\begin{figure}
\resizebox{\hsize}{!} {\includegraphics{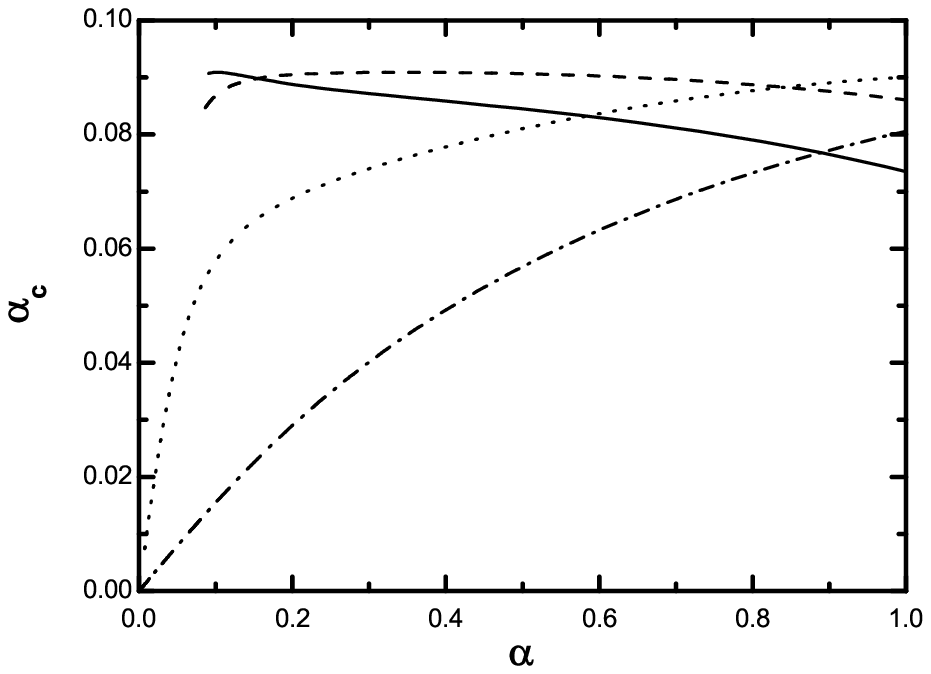}
\includegraphics{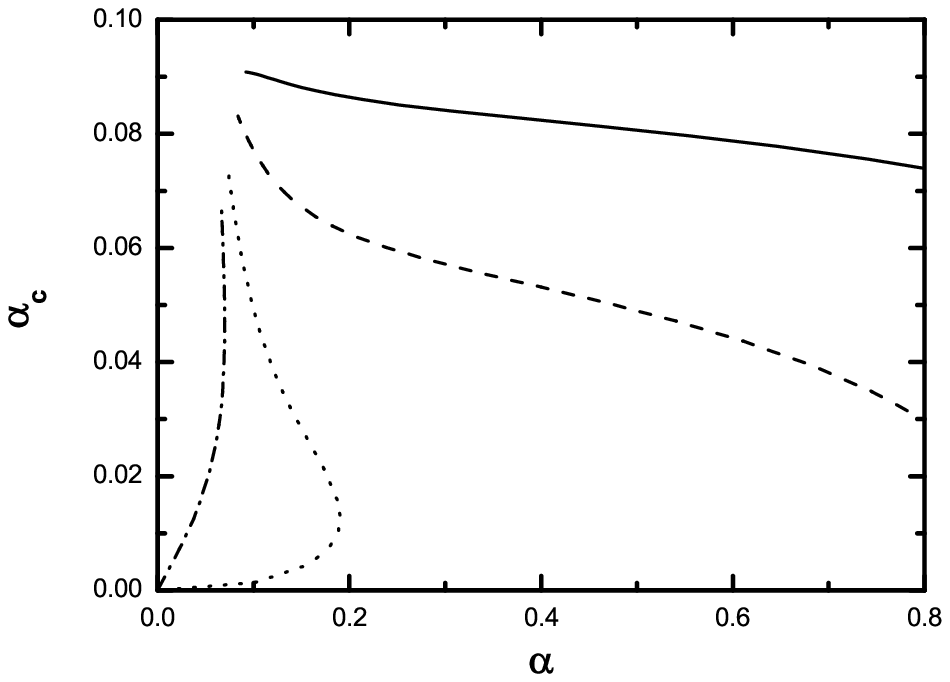}}
\caption{The convective coefficient $\alpha_{c}$ as a function of
viscosity parameter $\alpha$ with $s=-1/2$, $\gamma$=1.4, $f=1$ and
different sets of parameters ($\beta_{r}$,
$\beta_{\varphi}$,$\beta_{z}$). (a) {\em Left panel}: ($\beta_{r}$,
$\beta_{\varphi}$,$\beta_{z}$)$=(0, 0, 0)$ ({\em solid line}), $(0,
1, 0)$ ({\em dashed line}), $(0, 3, 0)$ ({\em dotted line}) and $(0,
5, 0)$ ({\em dash-dotted line}); (b) {\em Right panel}:
($\beta_{r}$, $\beta_{\varphi}$,$\beta_{z}$)$=(0, 0, 0.1)$ ({\em
solid line}), $(0, 0, 0.5)$ ({\em dashed line}), $(0, 0, 0.7)$ ({\em
dotted line}) and $(0, 0, 0.8)$ ({\em dash-dotted line})}
\end{figure}

\newpage
\begin{figure}
\resizebox{\hsize}{!} {\includegraphics{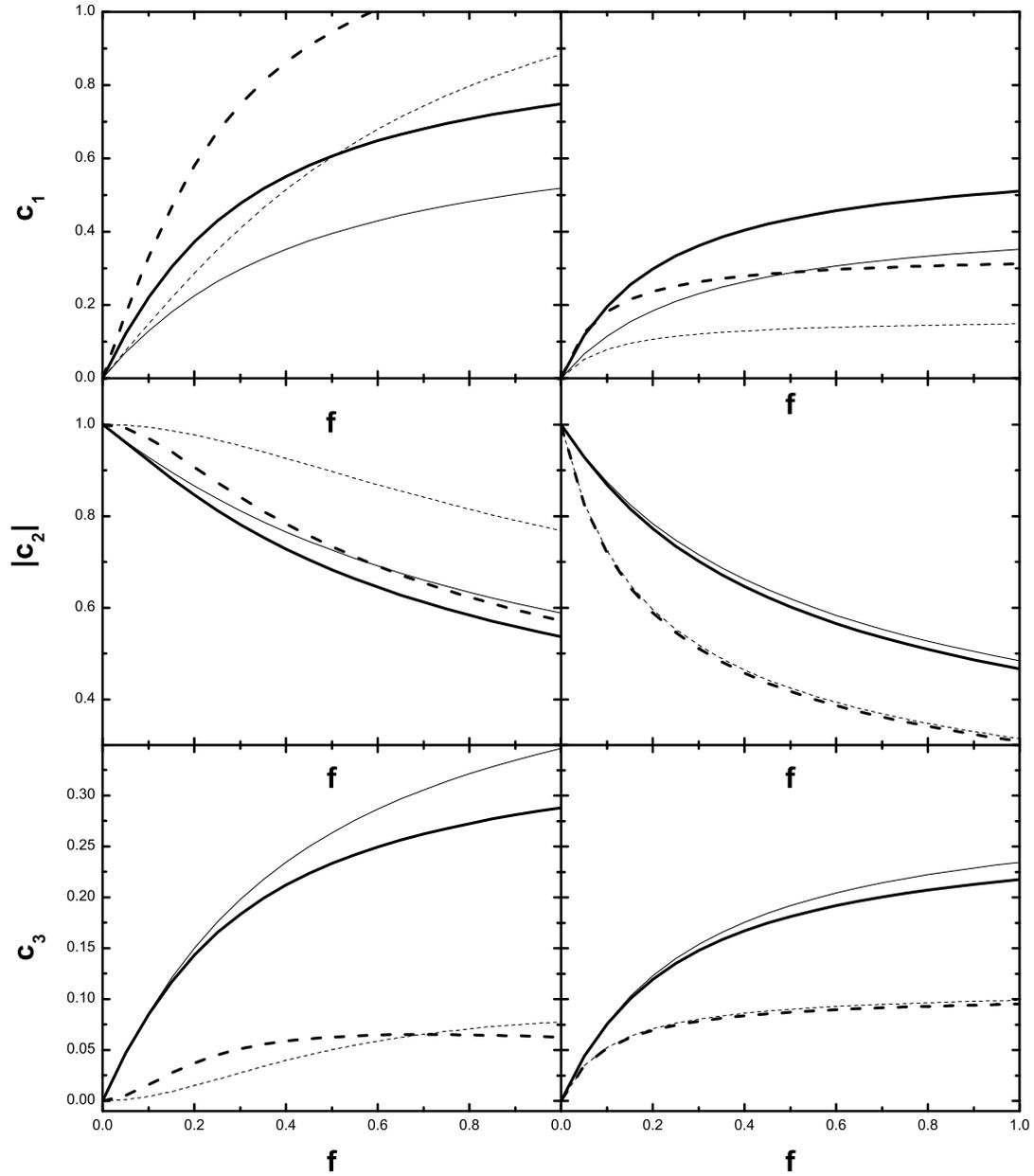}}
\caption{Comparison between two forms of the kinematic viscosity
$\nu$, while the results of $\nu=\alpha c_{s}^{2}/\Omega_{K}$ are
shown by thin lines, and those of $\nu=\alpha c_{s}H$ are shown by
thick lines. We adopt $s=-1/2$, $\gamma=4/3$ and $\alpha=1$. (a)
{\em Left panels}: ($\beta_{r}$, $\beta_{\varphi}$ $\beta_{z}$)$=(2,
0, 0)$ ({\em solid lines}) and $(2, 2.5, 0)$ ({\em dashed lines});
(b) {\em Right panels}: ($\beta_{r}$, $\beta_{\varphi}$
$\beta_{z}$)$=(2, 0, 1)$ ({\em solid lines}) and $(2, 0, 5)$ ({\em
dashed lines}). }
\end{figure}

\end{document}